\documentclass{pasj01}
\usepackage{natbib}
\usepackage{nidanfloat}

\begin{document} 
\Received{}
\Accepted{}


\title{MIRIS observation of near-infrared diffuse Galactic light}

\author{Y. \textsc{Onishi}\altaffilmark{1,5}}%
\author{K. \textsc{Sano}\altaffilmark{2,*}}
\author{S. \textsc{Matsuura}\altaffilmark{2}}
\author{W.-S. \textsc{Jeong}\altaffilmark{3,4}}
\author{J. \textsc{Pyo}\altaffilmark{3}}
\author{I.-J. \textsc{Kim}\altaffilmark{3}}
\author{H.~J. \textsc{Seo}\altaffilmark{3}}
\author{W. \textsc{Han}\altaffilmark{3,4}}
\author{D.-H. \textsc{Lee}\altaffilmark{3}}
\author{B. \textsc{Moon}\altaffilmark{3}}
\author{W.-K. \textsc{Park}\altaffilmark{3}}
\author{Y. \textsc{Park}\altaffilmark{3}}
\author{M.~G. \textsc{Kim}\altaffilmark{3}}
\author{T. \textsc{Matsumoto}\altaffilmark{3,5}}
\author{H. \textsc{Matsuhara}\altaffilmark{1,5}}
\author{T. \textsc{Nakagawa}\altaffilmark{5}}
\author{K. \textsc{Tsumura}\altaffilmark{7}}
\author{M. \textsc{Shirahata}\altaffilmark{5}}
\author{T. \textsc{Arai}\altaffilmark{5}}
\author{N. \textsc{Ienaka}\altaffilmark{6}}
\email{sano0410@kwansei.ac.jp}

\altaffiltext{1}{Department of Physics, Tokyo Institute of Technology, 2-12-1 Ookayama, Meguro-ku, Tokyo, 152-8550, Japan}
\altaffiltext{2}{Department of Physics, Kwansei Gakuin University, 2-1 Gakuen, Sanda, Hyogo 669-1337, Japan}
\altaffiltext{3}{Korea Astronomy and Space Science Institute, 776, Daedeok-daero, Yuseong-gu, Daejeon 305-348, Republic of Korea}
\altaffiltext{4}{Korea University of Science and Technology, 217, Gajeong-ro, Yuseong-gu, Daejeon 34113, Korea}
\altaffiltext{5}{Institute of Space and Astronautical Science, Japan Aerospace Exploration Agency, 3-1-1 Yoshinodai, Chuo-ku, Sagamihara, Kanagawa 252-5210, Japan}
\altaffiltext{6}{Department of Astronomy, Graduate School of Science, The University of Tokyo, Hongo 7-3-1, Bunkyo-ku, Tokyo 113-0033, Japan}
\altaffiltext{7}{Frontier Research Institute for Interdisciplinary Science, Tohoku University, Sendai 980-8578, Japan}




\maketitle

\begin{abstract}
We report near-infrared (IR) observations of high Galactic latitude  clouds to investigate diffuse Galactic light (DGL), which is starlight scattered by interstellar dust grains. 
The observations were performed at $1.1$ and $1.6\,\rm{\mu m}$ with a wide-field camera instrument, the Multi-purpose Infra-Red Imaging System (MIRIS) onboard the Korean satellite STSAT-3. 
The DGL brightness is measured by correlating the near-IR images with a far-IR $100\,\rm{\mu m}$ map of interstellar dust thermal emission. 
The wide-field observation of DGL provides the most accurate DGL measurement achieved to date. 
We also find a linear correlation between optical and near-IR DGL in the MBM32 field.
To study interstellar dust properties in MBM32, we adopt recent dust models with or without $\rm{\mu m}$-sized very large grains and predict the DGL spectra, taking into account reddening effect of interstellar radiation field.
The result shows that observed color of the near-IR DGL is closer to the model spectra without very large grains.
This may imply that dust growth in the observed MBM32 field is not active owing to its low density of interstellar medium.
\end{abstract}

\section{Introduction}

Diffuse Galactic light (DGL) is scattered light by interstellar dust grains illuminated by interstellar radiation field (ISRF). 
The DGL measurement is important in constraining interstellar dust properties because the spectrum reflects albedo, composition, and size distribution of grains.
In addition, the DGL is one of foreground emissions of the extragalactic background light (EBL).
Therefore, the DGL component should be subtracted accurately in the EBL measurement (e.g., Matsuoka et al. 2011; Matsumoto et al. 2015; Sano et al. 2015; Matsuura et al. 2017; Mattila et al. 2017).

The ISRF is not only scattered but also absorbed by interstellar dust and is re-emitted in the far-infrared (IR). 
Thus, the DGL is expected to correlate with the far-IR thermal emission since both components are related to column density of interstellar dust. 
In the optically thin case, this correlation should be linear, as studied by Brandt \& Draine (2012) and Ienaka et al. (2013).  
Based on the correlation analysis, a number of studies have measured the DGL in various regions, such as high-latitude general fields, diffuse clouds, and reflection nebulae (e.g., Witt et al. 2008; Brandt \& Draine 2012; Arai et al. 2015; Kawara et al. 2017).

According to scattering properties inferred from the Mie theory, scattering anisotropy depends on grain size ($a$) and wavelength ($\lambda$) of incident light.
Due to the anisotropic scattering of interstellar dust illuminated by  starlight from the Galactic plane, intensity ratio of the DGL to $100\,\rm{\mu m}$ emission changes as a function of Galactic latitude $b$ (Jura 1979; Sano et al. 2016b; Sano \& Matsuura 2017). 
This indicates that DGL observation toward identical Galactic latitudes is useful to study interstellar dust properties without considering the $b$-dependence.

Recent interstellar dust models comprise silicate and graphite grains with large molecules, such as polycyclic aromatic hydrocarbon (PAH).
They show continuous size distribution with grain size $a$ ranging from $\rm{n m}$ to sub-$\rm{\mu m}$.
Numerous dust models have been developed so far, including Weingartner \& Draine (2001, hereafter WD01), Zubko et al. (2004, hereafter ZDA04), and Jones et al. (2013).
These models marginally succeed in reproducing the observed interstellar extinction curve in most of the wavelengths (e.g., Cardelli, Clayton, \& Mathis 1989).
In the wavelengths $\lambda \sim 3$--$8\,\rm{\mu m}$, however, the models assuming Galactic average $R_V = 3.1$ are inconsistent with flat extinction curve observed by several studies (Lutz 1999; Indebetouw et al. 2005; Jiang et al. 2006; Flaherty et al. 2007; Gao et al. 2009; Nishiyama et al. 2009; Wang et al. 2013).
In addition, some studies have reported unexpectedly high albedo of interstellar dust in $\lambda \sim 2\,\rm{\mu m}$ (Witt et al. 1994; Block 1996; Lehtinen \& Mattila 1996), while models of Draine \& Lee (1984) and Li \& Greenberg (1997) predict that the albedo at $2.2\,\rm{\mu m}$ is about $0.2$.
These results may suggest presence of dust grains larger than sub-$\rm{\mu m}$, referred to as very large grains (VLGs), because such populations are expected to enhance the near- to mid-IR scattering efficiency according to the Mie theory.

Wang, Li, \& Jiang (2015a; hereafter WLJ15) suggests a new dust model including graphite grains larger than $\sim 1\,\rm{\mu m}$.
This model explains the mid-IR flat extinction curve and is also consistent with dust thermal emission observed with Diffuse Infrared Background Experiment (DIRBE; Arendt et al. 1998), Far Infrared Absolute Spectrophotometer (FIRAS; Finkbeiner et al. 1999), and {\it Planck} (Planck Collaboration XVII 2014).
Though the VLG population may also change grain albedo and affect the DGL spectrum, Ienaka et al. (2013) points out that the observed DGL results in various regions differ by a factor of 2 due to the $b$-dependence from the anisotropic scattering and possible statistical uncertainty.
To investigate the VLG population from the DGL observation, more precise measurement of the DGL is helpful.

In this study, we present DGL measurement with the best accuracy to date in the near-IR photometric bands by using the wide-field imaging capability of Multi-purpose Infra-Red Imaging System (MIRIS; Han et al. 2014). 
We also derive color of the DGL from optical to near-IR, which is useful for constraining grain size distribution of interstellar dust.
Then, we compare the present DGL observation with the model spectra expected from recent dust models with or without VLGs.

Remainder of this paper is organized as follows.
In Section 2, we describe the MIRIS instrument and observation in the present study.
In Section 3, we present analysis of the observed images to measure the near-IR DGL.
Section 4 shows the results of correlation analysis of the near-IR DGL against the $100\,\rm{\mu m}$ emission and optical DGL.
In Section 5, we discuss size distribution of interstellar dust on the basis of dust models with or without VLG.
Summary appears in Section 6.

\section{Observations}

\subsection{MIRIS}

MIRIS was developed by the Korean Astronomy and Space Science Institute (KASI) and was launched onboard Science and Technology Satellite-3 (STSAT-3) in November 2013. 
Primary purposes of MIRIS are survey of Paschen-$\alpha$ emission lines from the Galactic plane and detection of EBL in high Galactic latitudes.
The optical design of MIRIS was optimized for a wide-field survey in near-IR wavelengths.

MIRIS has two broad $I$- and $H$-band, hereafter referred to as $1.1$ and $1.6\,\rm{\mu m}$ bands, respectively. 
MIRIS also consists of two narrow-bands comprising the Paschen-$\alpha$ line band ($1.876\,\rm{\mu m}$) and the dual-band Continuum ($1.84$ and $1.91\,\rm{\mu m}$).
MIRIS adopted $256\times 256$ PICNIC HgCdTe sensor array of a $3.67^{\circ} \times 3.67^{\circ}$ field of view (FOV) with a pixel scale of $51.6\,\rm{arcsec}$.
See Han et al. (2014) for detail about the MIRIS instruments.

\subsection{Field selection}

\begin{table*}
 \caption{Information of three fields observed with MIRIS}
  \label{table:table1}
  \centering
  \begin{tabular}{ccccc}
    \hline
     Field name   &   Date of the observation   &  Ecliptic coordinates $\left( \lambda ,\beta  \right) $ [deg]   &  Galactic coordinates $\left( l,b  \right) $ [deg]   &  Max $I_{100\,\rm{\mu m}}$ [$\rm{MJy\,sr^{-1}}$] \\
    \hline \hline
    CIBDC3 & Jan 13, 2015 & (41.01, 2.04) & (154.60, -39.70) & $14.69$ \\
    MBM32 & Feb 25, 2015  & (119.06, 47.61) & (147.16, 40.62) & $7.95$ \\
   CIBDC5 & Mar 20, 2015  & (115.17, 52.49) & (140.98, 38.21) & $10.01$ \\
    \hline
  \end{tabular}
  \medskip
\end{table*}

To achieve precise correlation analysis between near-IR DGL and far-IR dust emission, we need to observe diffuse clouds showing wide intensity range at $100\,\rm{\mu m}$.
In addition, the clouds should be located in high-$b$ regions to reduce contamination of Galactic stars. 
To select appropriate clouds, we used optical images of Digital Sky Survey and $100\,\rm{\mu m}$ intensity map created by Schlegel, Finkbeiner, and Davis (1998; hereafter SFD98) on the basis of all-sky observation of DIRBE on board {\it Cosmic Background Explorer} ({\it COBE}; Hauser et al. 1998).
Then, we found two fields, namely CIBDC3 and CIBDC5. 
To measure the DGL color between the near-IR and optical in an identical Galactic latitude, we also chose MBM32 cloud, where the optical observation was already conducted by Ienaka et al. (2013). 
Using MIRIS, we observed these targets at $1.1$ and $1.6\,\rm{\mu m}$ from January to March 2015.
Properties of the three fields are shown in Table \ref{table:table1}.

\section{Analysis}

\subsection{Data reduction}

Observed images are processed by data reduction pipeline established by KASI (Han et al. 2014). 
One observation of MIRIS consists of $360$ frames including $120$ dark and $240$ observations, where $1$ frame corresponds to $2$-sec exposure.
Frames are reset once every 10 frames.
 
\begin{figure*}
\begin{center}
  \includegraphics[width=6cm,angle=90]{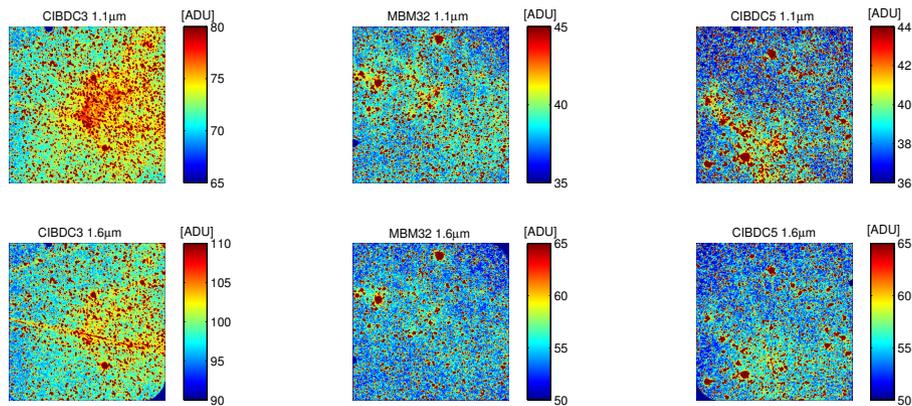}
  \caption{Images of our three fields, CIBDC3, MBM32, and CIBDC5 observed by MIRIS at $1.1$ and $1.6\,\rm{\mu m}$, calibrated to values in units of ADU.
  Each field is approximately $3.67^{\circ} \times 3.67^{\circ}$.
  }
  \label{fig:figure1}
\end{center}
\end{figure*}

Processing of the observed images includes following processes: bad pixel masking, linearity correction of detector signals, frame differentiation, flat-field correction, distortion correction, astrometry correction, and stacking frames.
In the linearity correction, saturated pixels are removed and measured pixel values are divided with forth order polynominal function.
In the differentiation, frame-to-frame difference of pixel values are calculated.
Flat template of MIRIS is adopted in the flat-field correction.
The Distortion correction is conducted by the IRAF GEOTRAN task.
After astrometry correction of the image, the differentiated frames were stacked to create the FITS data.
Images of the three clouds after these processes are shown in Figure \ref{fig:figure1} in units of Analog to Digital Unit (ADU).

\subsection{Flux calibration}

Standard method for flux calibration of the MIRIS detector has not yet been developed by the KASI team.
Thus, we carry it out by using stars in our images.
First, we select $50$ stars in our images, whose spectral types are known from Two Micron All-Sky Survey (2MASS) Point Source Catalog (PSC; Cutri et al. 2003; Skrutskie et al. 2006). 
This is because we need spectral color correction for very broad profiles of the MIRIS filters against the 2MASS bands.
As shown in the left panel of Figure 2, we stack these stars in the image to estimate point spread function (PSF) on the detector array. 
We approximate the beam profile as a two-dimensional Gaussian. 
Estimated variance along $x$ and $y$ directions are $\sigma_x = 1.14 \pm 0.04$ and $\sigma_y = 1.04 \pm 0.10$ pixel, corresponding to $0.98 \pm 0.03$ and $0.89 \pm 0.09\,\rm{arcmin}$, respectively. 
Results of Gaussian fitting along the $x$ and $y$ directions are shown in the right panel of Figure 2.

\begin{figure*}
\begin{center}
  \includegraphics[width=10cm,clip=true]{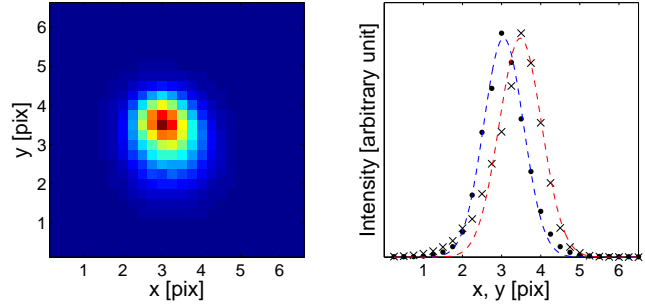}
  \caption{Left panel: stacked image of the selected 2MASS stars on the MIRIS detector at $1.1\,\rm{\mu m}$. 
  Right panel: crosscut profiles of the beam (left panel) along $x$ and $y$ directions.
  Blue and red dashed curves are Gaussian functions along $x$ and $y$ directions, respectively, estimated from the two-dimensional fitting to the staked image.} 
  \label{fig:figure31}
  \end{center} 
\end{figure*}

Next process is to determine spectral radiance of the selected stars. 
We assume that spectra of the stars are the blackbody functions whose temperature is estimated from the spectral types.  
At $1.1\,\rm{\mu m}$, we adopt the 2MASS $J$-band flux as a scaling factor of the blackbody spectrum. 
At $1.6\,\rm{\mu m}$, 2MASS $H$-band flux is used. 
Adopting the scaling factor $S_i$ of star ID $i$, we can express the star flux $F_i$ as
\begin{equation}
F_i = \frac{S_i\int_{0}^{\infty}{B_i\left(\lambda  \right) T\left(\lambda  \right) d\lambda}}{\int_{0}^{\infty}{ T\left( \lambda  \right) d\lambda}},
\end{equation}
where $T(\lambda)$ is transmittance of the MIRIS filters at $1.1$ or $1.6\,\rm{\mu m}$ as shown in Figure 3.  
The blackbody spectrum of star ID $i$ is expressed as $B_i(\lambda)$. 
Then, we obtain flux calibration factor $C_i$ of star ID $i$ as
\begin{equation}
C_i=\frac{\lambda_{C}{F_i}}{\Omega_p \iint{P\left(x,y \right) dxdy}}, 
\end{equation}
in units of ${\rm nW\,m^{-2}\,sr^{-1}}\,{{\rm ADU}^{-1}}$, where $P\left(x,y \right)$ is the beam profile, $\Omega_p$ is FOV of a pixel in unit of sr, $\lambda_C$ is the central wavelength of the two bands, $1.076$ and $1.608\,\rm{\mu m}$. 

We present the flux calibration factor $C_i$ of each star in Figure 4. 
Because of incompleteness in the linearity correction, a few bright stars of $\sim 0.15\,{\rm nW\,m^{-2}\,sr^{-1}}$ at $1.1\,\rm{\mu m}$ show exceptionally high values of $C_i$. 
We exclude these bright stars and estimate the mean values of the calibration factor as $11.83 \pm 0.34$ and $6.02 \pm 0.11\,{\rm nW\,m^{-2}\,sr^{-1}}\,{{\rm ADU}^{-1}}$ at $1.1$ and $1.6\,\rm{\mu m}$, respectively. 
These values include both statistical uncertainty of $C_i$ and photometric errors of the standard stars. 

\begin{figure*}
\centering
 \includegraphics[width=10cm,clip=true]{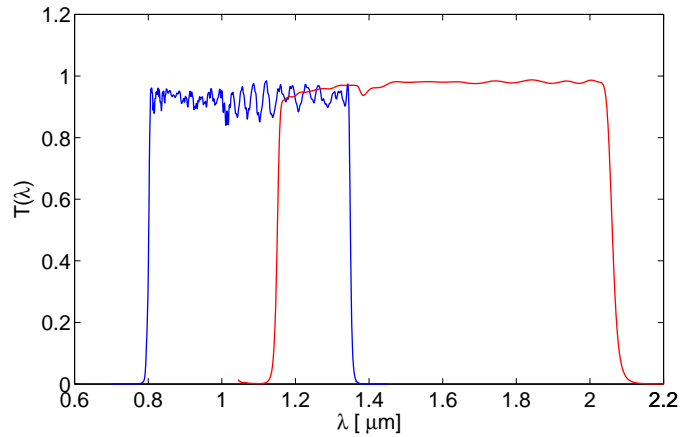}
 \caption{Relative transmittance of MIRIS photometric bands at $1.1\,\rm{\mu m}$ (blue) and $1.6\,\rm{\mu m}$ (red).}
 \label{fig:figure32}
\end{figure*}

Sensitivity of the MIRIS detector changes slightly during its observations. 
As it is difficult to correct this effect by modeling the sensitivity changes, we instead incorporate the time variance as a systematic error of the calibration factor.
As root mean square of the variance in the three fields, we derive the systematic uncertainties of $1.9\%$ and $2.4\%$ at $1.1$ and $1.6\,\rm{\mu m}$, respectively. 
Finally, we estimate the total calibration factor of $11.83 \pm 0.56$ and $6.02 \pm 0.27 \,{\rm nW\,m^{-2}\,sr^{-1}}\,{{\rm ADU}^{-1}}$ at $1.1$ and $1.6\,\rm{\mu m}$, respectively.

\begin{figure*}
\centering
 \includegraphics[width=8cm,clip=true]{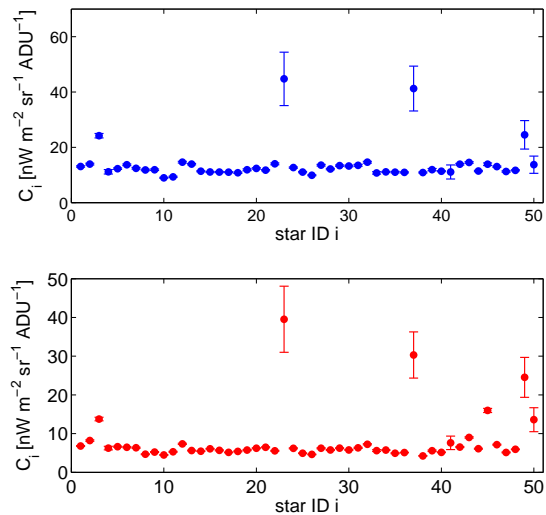}
 \caption{Flux calibration factor $C_i$ of star ID $i$. 
 Top and bottom panels indicate flux calibration factors at $1.1$ and $1.6\,\rm{\mu m}$, respectively.
 Each error bar is derived from photometric uncertainty provided by 2MASS PSC.
 A few exceptional values are excluded in calculating the averaged calibration factor (Section 3.2).}
 \label{fig:figure33}
\end{figure*}

\subsection {Point source masking}

To analyze diffuse emissions, it is necessary to remove bright point sources from the image. 
At first, we establish compartments including $\sim 10 \times 10$ pixels in all parts of the images, which are larger than the MIRIS PSF but are smaller than the cloud size.  
Then, we calculate the means and standard deviation of the values in each compartment and continue to perform $2\sigma$ clipping until the clipped pixels disappear.
After that, means plus the $2\sigma$ value are assumed as sky noise and are assigned to the central pixel of each compartment.
In this way, images of the sky noise are created.

Second, we generate integrated brightness map of the stars by convolving the MIRIS PSF with the 2MASS sources on the obtained images.
The 2MASS PSC contains photometry of the objects covering $99.998\%$ of the sky with accurate detections below the completeness limit of $J = 15.8$ and $H = 15.1\,{\rm mag}$ in Vega photometric system (Cutri et al. 2003; Skrutskie et al. 2006). 
We use the point sources in the 2MASS $H$-band to create the integrated brightness map.  
To remove contribution of stars, pixels whose values are higher than the sky noise are masked in the map.

In addition to the above processes, there are slight incompleteness of the masking of bright sources because of pointing instability of the MIRIS, which is approximately one pixel ($\sim {1}^{\prime}$) during the observations. 
Thus, we carry out the $2\sigma$ clipping method again. 
In this process, we calculate the means and $2\sigma$ values in each compartment and mask the pixels with larger or smaller values than the $2\sigma$ level. 
We iterate this procedure until clipped pixels disappear.

\subsection {Removal of the ZL gradient}

\begin{figure*}
\centering
\includegraphics[width=10cm]{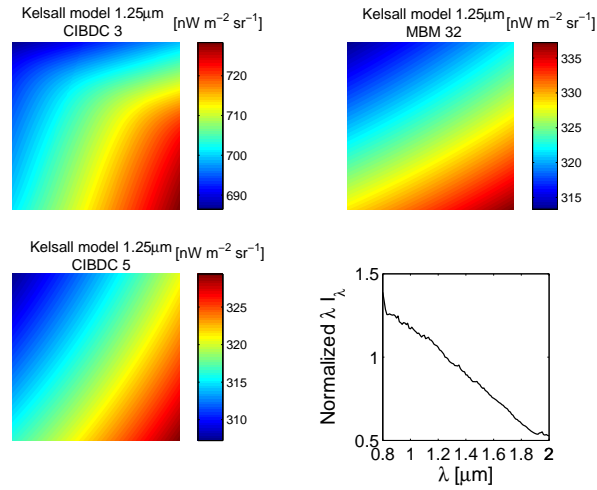}
 \caption{Gradient of the ZL brightness in the model of Kelsall et al. (1998) at $1.25\,\rm{\mu m}$ in each field. 
Image size is the same as Figure 1. 
The bottom right panel indicates the ZL spectrum observed with CIBER/LRS, normalized to unity at $1.25\,\rm{\mu m}$ (Arai et al. 2015).}
 \label{fig:figure35} 
\end{figure*}

Another necessary thing is removal of the zodiacal light (ZL).
The ZL is scattered sunlight by interplanetary dust and it dominates the near-IR sky brightness by more than $\sim80\%$.  
The ZL is uniform in spatial scales smaller than a degree (Pyo et al. 2012). 
However, we must remove the spatial gradient of the ZL because of the large FOV of MIRIS.

To remove the ZL gradient, we adopt a ZL model on the basis of the DIRBE observation (Kelsall et al. 1998). 
In the model, the ZL intensity is calculated in the DIRBE two photometric bands at $1.25$ and $2.2\,\rm{\mu m}$. 
Therefore, we need spectral information of ZL to estimate the ZL intensity in the MIRIS bands at $1.1$ and $1.6\,\rm{\mu m}$.
As a spectral template of ZL, we adopt the low resolution spectrum observed with Cosmic Infrared Background Experiment (CIBER) Low Resolution Spectrometer  (LRS; Tsumura et al. 2010; Arai et al. 2015). 
Using the model and spectrum of ZL, we estimate the ZL brightness in the MIRIS bands.
As shown in Figure 5, relative change of the ZL brightness within each field is an order of $\pm 3\%$. 
We subtract the ZL contribution from the total sky brightness to extract the DGL component.

\subsection{Smoothing of the MIRIS images}

\begin{figure*}
\centering
\includegraphics[width=12cm]{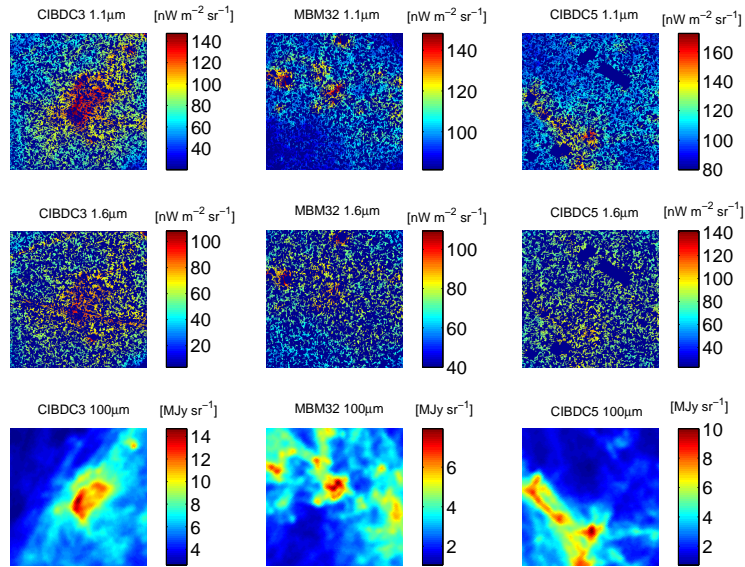}
\caption{Images of our targets after the reduction described in Section 3.
Top and middle panels indicate the images at $1.1$ and $1.6\,\rm{\mu m}$, respectively, calibrated to intensity units of ${\rm nW\,m^{-2}\,sr^{-1}}$.
Bottom panels are the SFD98 $100\,\rm{\mu m}$ map of each field in units of $\rm{MJy\,sr^{-1}}$.
Image size is the same as Figure 1. 
}
\label{fig:figure36} 
\end{figure*}

We use the SFD98 $100\,\rm{\mu m}$ map for correlation analysis against the near-IR DGL.
Spatial resolution of the SFD98 map is approximately $2_\cdot^{\prime}372  \times 2_\cdot^{\prime}372$ and is lower than that of MIRIS ($\sim {0}_{{\large{\cdot}}}^{\prime}9 \times {0}_{{\large{\cdot}}}^{\prime}9$). 
To compare the MIRIS data with the $100\,\rm{\mu m}$ map, we reduce the resolution of the near-IR images to that of the SFD98 map by averaging $3 \times 3$ pixels of the MIRIS images. 
In Figure \ref{fig:figure36}, we present the smoothed near-IR images in comparison with the SFD98 $100\,\rm{\mu m}$ map.

\section{Results}

\subsection{Correlation between DGL and $100\,\rm{\mu m}$ emission}

\begin{figure*}
\centering
\includegraphics[width=8cm,clip=true]{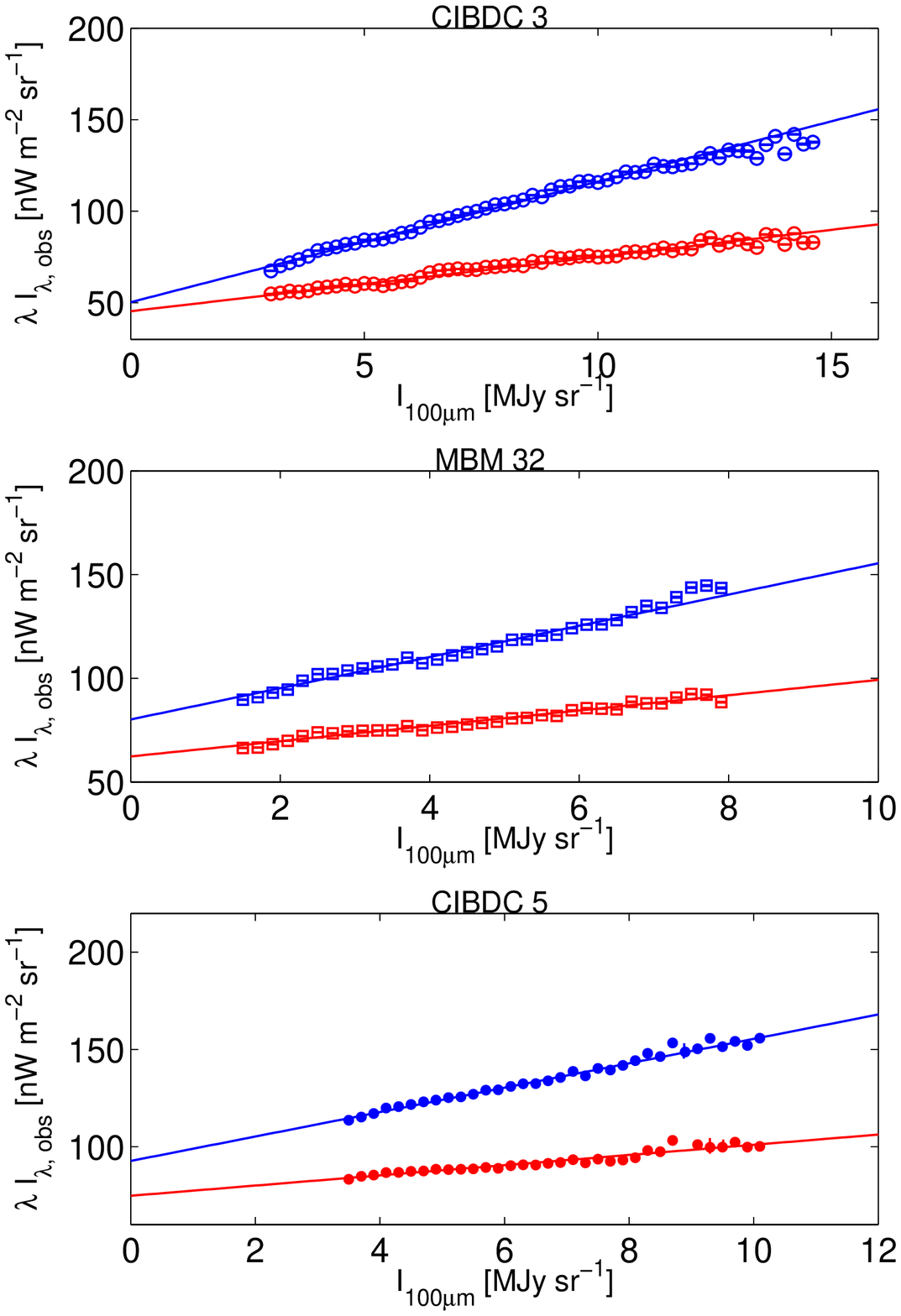}
\caption{Correlation between the near-IR MIRIS images and SFD98 $100\,\rm{\mu m}$ map. 
Top, middle, and bottom panels indicate the correlation in the fields of CIBDC3, MBM32, and CIBDC5, respectively. 
Blue and red points indicate, respectively, the average and standard errors of $\lambda I_{\lambda, {\rm obs}}$ at $1.1$ and $1.6\,\rm{\mu m}$ in a certain bin of $I_{100\,\rm{\mu m}}$.
The blue solid lines in each panel represent best-fit lines of the correlation between $1.1\,\rm{\mu m}$ and $100\,\rm{\mu m}$, while the red solid lines are those between $1.6\,\rm{\mu m}$ and $100\,\rm{\mu m}$. 
}
\label{fig:figure6} 
\end{figure*}

Prior to correlating the near-IR images with the SFD98 map, we clarify the relation between DGL and far-IR emission. 
According to Henyey (1937), DGL brightness $I_{\lambda, {\rm DGL}}$ through a dust slab is written as
\begin{equation}
I_{\lambda, {\rm DGL}} = \frac{\gamma_\lambda}{1-\gamma_\lambda} I_{\lambda, {\rm ISRF}} \left[1-e^{-\left(1-\gamma_\lambda \right) \tau_{\lambda,{\rm ext}}} \right], 
\end{equation}
where $\gamma_\lambda$ and $\tau_{\lambda,{\rm ext}}$ represent, respectively, the albedo and optical depth for extinction.
The ISRF brightness is represented as $I_{\lambda, {\rm ISRF}}$. 
This formula does not include the $b$-dependence caused by the scattering anisotropy of dust grains (e.g., Jura 1979).
In optically thin limit $\tau_{\lambda,{\rm ext}} \ll 1$, Equation (3) is approximated as 
\begin{equation}
I_{\lambda, {\rm DGL}} \approx \gamma_\lambda \tau_{\lambda,{\rm ext}} I_{\lambda, {\rm ISRF}}. 
\end{equation}
On the other hand, the brightness of far-IR $100\,\rm{\mu m}$ emission $I_{100\,\rm{\mu m}}$ is approximated as
\begin{equation}
I_{100\,\rm{\mu m}} \approx \left( 1-\gamma_{100\,\rm{\mu m}} \right) \tau_{100\,\rm{\mu m}, \rm{ext}} B_{100\,\rm{\mu m}}\left(T_D \right), 
\end{equation}
where $B_{100\,\rm{\mu m}}(T_D)$ indicates the Planck function of dust temperature $T_D$. 
The albedo $\gamma_{100\,\rm{\mu m}}$ should be nearly zero.
Then, the relation between $I_{\lambda, {\rm DGL}}$ and $I_{100\,\rm{\mu m}}$ are
\begin{equation}
I_{\lambda, {\rm DGL}} \approx \frac{\gamma_\lambda \tau_{\lambda,{\rm ext}} I_{\lambda, {\rm ISRF}}}{\left(1-\gamma_{100\,\rm{\mu m}} \right) \tau_{100\,\rm{\mu m}, {\rm ext}}B_{100\,\rm{\mu m}}\left(T_D \right)} I_{100\,\rm{\mu m}}.
\end{equation}
From Equation (6), a linear correlation between the brightness of DGL and that of the $100\,\rm{\mu m}$ emission is expected in the optically thin case.

In addition to the DGL, observed diffuse light includes isotropic EBL and the integrated very faint stars. 
Therefore, we approximate the near-IR sky brightness $I_{\lambda, {\rm obs}}$ as
\begin{equation}
I_{\lambda, {\rm obs}}=a_\lambda+b_\lambda I_{100\,\rm{\mu m}},
\end{equation}
where the slope $b_\lambda$ indicates the intensity ratio of DGL to $100\,\rm{\mu m}$ emission, while $a_\lambda$ indicates contribution of the isotropic components.

Figure 7 shows correlation between the near-IR DGL and SFD98 $100\,\rm{\mu m}$ emission along with best-fit lines by Equation (7).
Each point indicates the average with the standard errors in a certain bin of $I_{100\,\rm{\mu m}}$, though most of the error bars are small. 
All clearly show linear correlations against the $100\,\rm{\mu m}$ brightness.

The parameters $a_\lambda$ and $b_\lambda$ derived in each field are summarized in Table \ref{table:table2}. 
In CIBDC3, the parameter $a_\lambda$ is significantly smaller than the values derived in MBM32 and CIBDC5. 
The difference is thought to originate from uncertainty in the ZL model (Section 3.4). 
As shown in Table \ref{table:table1}, CIBDC3 is located at a low ecliptic latitude, where the ZL is roughly twice as strong as it is for MBM32 and CIBDC5.
We also present weighted mean of $\lambda b_\lambda$ in the three field (Table 2).

Figure 8 compiles recent observations of the parameter $\lambda b_\lambda$ from optical to near-IR.
Our result smoothly connects to the DIRBE observation (Sano et al. 2015; 2016a) and is marginally consistent with the DGL spectrum derived from CIBER/LRS (Arai et al. 2015). 
Thanks to the large FOV and the broad-band observation of MIRIS, our results show the highest precision among the near-IR results.

\begin{table*}
 \caption{Summary of linear correlation parameters}
  \label{table:table2}
  \footnotesize
  \scalebox{0.5}
  \centering
  \begin{tabular}{ccccc} 
   \hline
     Field name   &   $\lambda a_{1.1\,\rm{\mu m}}$    &    $\lambda b_{1.1\,\rm{\mu m}}$  &  $\lambda a_{1.6\,\rm{\mu m}}$   &  $\lambda b_{1.6\,\rm{\mu m}}$\\
   &   $\rm{nW\,m^{-2}\,sr^{-1}}$ & $\rm{nW\,m^{-2}\,sr^{-1}/(MJy\,sr^{-1})}$ & $\rm{nW\,m^{-2}\,sr^{-1}}$ &  $\rm{nW\,m^{-2}\,sr^{-1}/(MJy\,sr^{-1})}$   \\
    \hline \hline
    CIBDC3  &   50.3 $\pm$ 3.1      &      6.23 $\pm$ 0.10      &      42.4 $\pm$ 2.8      &       2.51 $\pm$ 0.09 \\
    MBM32    & 84.5 $\pm$ 5.2      &      7.64 $\pm$ 0.32      &      60.5 $\pm$ 2.7      &      3.59 $\pm$ 0.27 \\
 CIBDC5      &      94.7 $\pm$ 5.3      &      6.56 $\pm$ 0.12      &      72.1 $\pm$ 3.2      &      2.63 $\pm$ 0.10 \\
    \hline
    Weighted mean   & ------ & 6.43 $\pm$ 0.87 &  ------  & 2.62 $\pm$ 0.69 \\
    \hline
  \end{tabular}
  \medskip

\end{table*}

\begin{figure*}
\centering
 \includegraphics[width=10cm,clip=true]{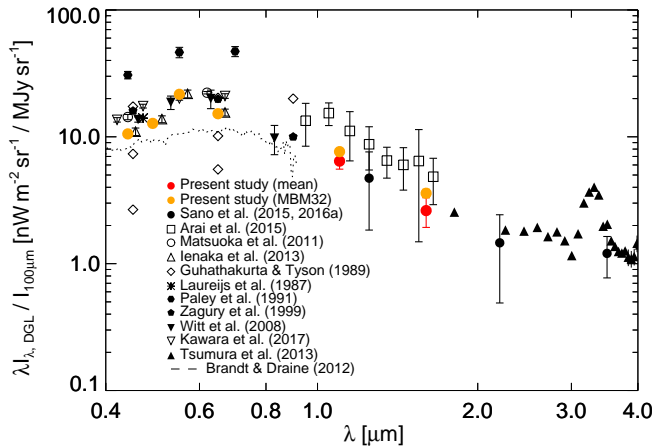}
 \caption{Intensity ratios of optical to near-IR DGL to $I_{100\,\rm{\mu m}}$, observed in different fields of the sky. 
 The red circles represent the weighted mean of the present results in the three fields (Table 2). 
  The orange circles in the near-IR indicate the present result in the MBM32 field, while those in the optical are from correlation between the near-IR and optical data (Figure 10). 
  Some points are shifted from the exact wavelengths for clarity.
 References of earlier results appear in the figure.}
 \label{fig:figure9}
\end{figure*}

\subsection{Correlation between optical and near-IR DGL}

To derive color of DGL over a wide range from optical to near-IR wavelengths, we correlate the present near-IR MBM32 image with the optical observations in the $B$, $g$, $V$, and $R$ bands (Ienaka et al. 2013). 
The optical data were acquired by the $105\,\rm{cm}$ Schmidt telescope with a $2048 \times 2048$ pixel 2KCCD camera at Kiso observatory in February and April 2011 and February 2012. 
The FOV of the camera is $50^{\prime} \times 50^{\prime}$ with a pixel scale of $1_\cdot^{\prime \prime}5$. 

To take correlation between optical and near-IR, we reduce spatial resolution of the optical images to that of MIRIS. 
The near-IR and optical images in the MBM32 field are compared in Figure 9. 
We take correlation between the MIRIS $1.1\,\rm{\mu m}$ and optical bands as shown in Figure 10. 
We find linear correlation between the optical and near-IR DGL for the first time.
Against the MIRIS data at $1.1\,\rm{\mu m}$, intensity ratios derived by linear fitting are $1.38\pm0.12$, $1.67\pm0.12$, $2.83\pm0.22$, and $1.99\pm0.13$ at $0.44$, $0.49$, $0.55$, and $0.65\,\rm{\mu m}$, respectively, in units of $\rm{nW\,m^{-2}\,sr^{-1}}/\rm{nW\,m^{-2}\,sr^{-1}}$.
The linear correlation may indicate that several tens of $\rm{nm}$ to sub-$\rm{\mu m}$-sized grains contributing to the DGL in the optical and near-IR are well mixed in the MBM32 field.

From the intensity ratios in Figure 10 and near-IR to $100\,\rm{\mu m}$ correlation (Figure 7), we calculate intensity ratios of the optical DGL to $100\,\rm{\mu m}$ emission.
The results are indicated by orange circles in Figure 8 and are nearly the same as those derived from direct linear correlation between the optical DGL and $100\,\rm{\mu m}$ emission (Ienaka et al. 2013).
This finding makes the correlation method against $100\,\rm{\mu m}$ emission more reliable and convincing.

\begin{figure*}
\centering
 \includegraphics[width=80mm,clip=true,bb=10   201   694   590]{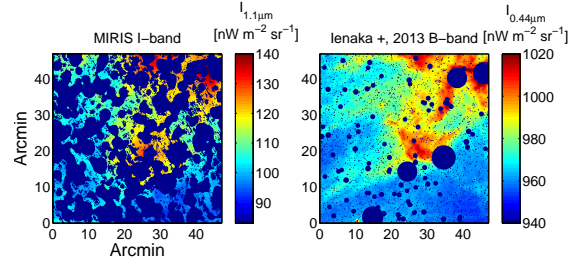}
 \caption{Images of the MBM32 field at $1.1$ and $0.44\,\rm{\mu m}$. 
 The left panel is the MIRIS image whose spatial scale is the same as the optical images in the right panel (Ienaka et al. 2013).  
 Sky brightness is in units of $\rm{nW\,m^{-2}\,sr^{-1}}$. }
 \label{fig:figure7} 
\end{figure*}

\begin{figure*}
\begin{center}
\centering
 \includegraphics[width=80mm,angle=90]{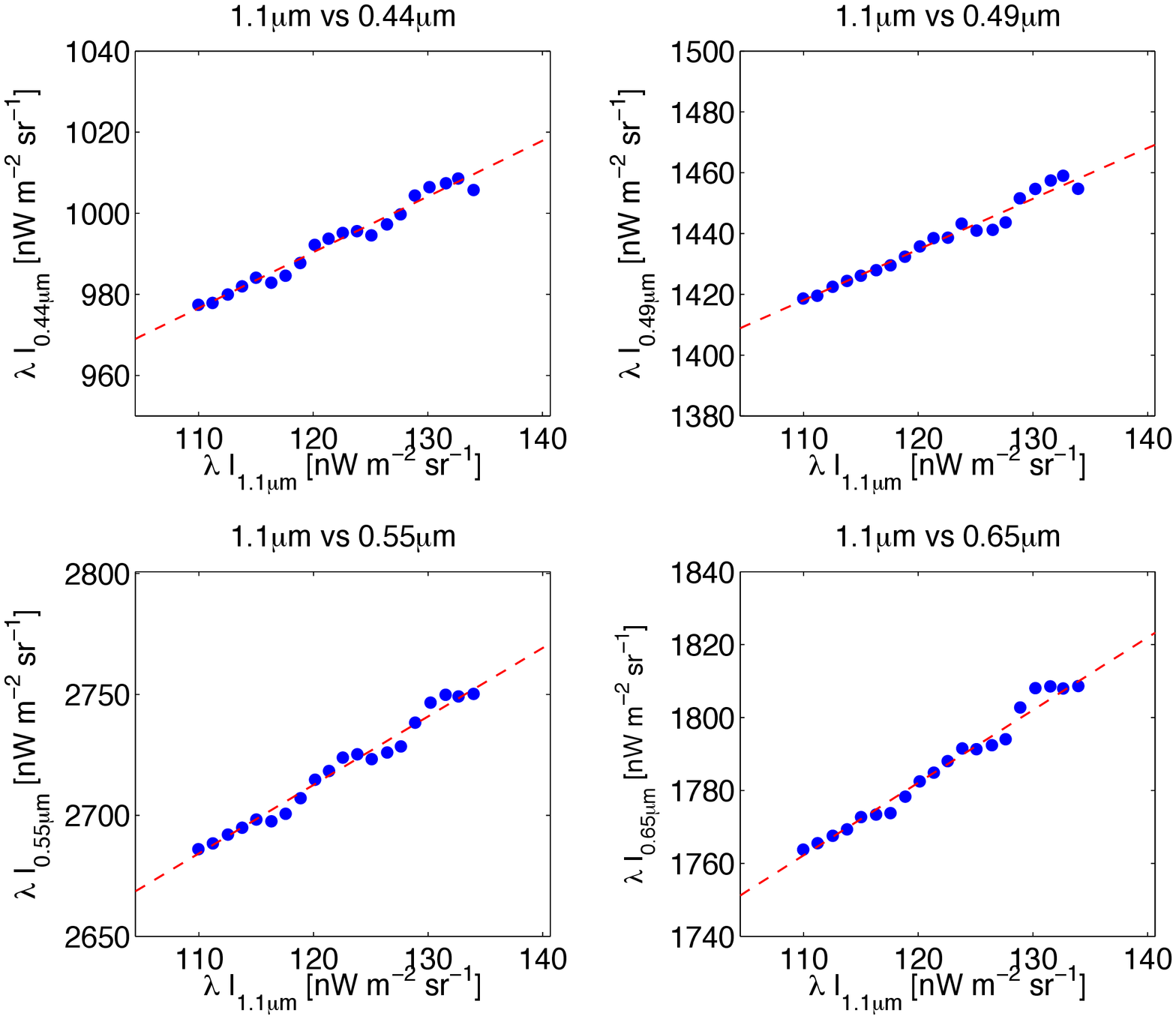}
 \caption{Correlation between the image at MIRIS $1.1\,\rm{\mu m}$ and those in the optical four bands (Ienaka et al. 2013). 
 In each panel, blue circles and red dashed line represent averaged optical intensity in a certain $x$-axis bin and the best-fit line, respectively. }
  \label{fig:figure8} 
 \end{center}
\end{figure*}

\section{Discussion}

Based on our precise DGL measurements from optical to near-IR, we investigate properties of interstellar dust.
The dust model including $\rm{\mu m}$-sized graphite reportedly reproduces the mid-IR flat extinction curve and the IR emission spectrum (WLJ15).
Though the increase of C/H abundance exceeds the solar metallicity (Asplund et al. 2009), it can be consistent with gas-phase C/H abundance observed toward several sightlines in the Milky Way (Parvathi et al. 2012).
However, it is unclear whether the WLJ15 model is also applicable to the DGL observation.
We therefore calculate the DGL spectrum from the WLJ15 model and compare it with the present observation.

Brandt \& Draine (2012) estimates the DGL spectrum in a plane-parallel galaxy model, assuming different interstellar dust models of WD01 and ZDA04.
The WD01 model includes more sub-$\rm{\mu m}$-sized grains than ZDA04 and it causes difference in the DGL color from optical to near-IR.
Therefore, our DGL observation in the optical and near-IR are useful to constrain typical grain size of interstellar dust.

\subsection{Formulation of DGL from interstellar dust properties}

To clarify relations between the DGL spectrum and fundamental scattering properties of dust grains, we summarize some relevant quantities prior to specific calculations.
Grain shape is assumed to be sphere, which can be treated by the classical Mie theory.
If efficiency factors of extinction, scattering, and absorption are defined as $Q_{\rm ext} (\lambda, a)$, $Q_{\rm sca} (\lambda, a)$, and $Q_{\rm abs} (\lambda, a)$, respectively, relations between these quantities are 
\begin{equation}
Q_{\rm ext} (\lambda, a) = Q_{\rm sca} (\lambda, a) + Q_{\rm abs} (\lambda, a),
\end{equation}
\begin{equation}
Q_{\rm ext} (\lambda, a) \equiv \frac{\sigma_{\rm ext} (\lambda)}{\pi a^2},
\end{equation}
\begin{equation}
Q_{\rm sca} (\lambda, a) \equiv \frac{\sigma_{\rm sca} (\lambda)}{\pi a^2},
\end{equation}
\begin{equation}
Q_{\rm abs} (\lambda, a) \equiv \frac{\sigma_{\rm abs} (\lambda)}{\pi a^2},
\end{equation}
where $\sigma_{\rm ext} (\lambda)$, $\sigma_{\rm sca} (\lambda)$, and $\sigma_{\rm abs} (\lambda)$ are cross sections of extinction, scattering, and absorption, respectively.
If number density of hydrogen and that of dust grains in the size range $a \sim a+da$ are defined as ${n_{\rm H}}$ and $dn/da$, respectively, interstellar extinction $A_{\lambda}$ per hydrogen column density $N_{\rm H}$ is calculated as
\begin{equation}
A_{\lambda}/N_{\rm H} = 1.086 \tau_{\lambda,{\rm ext}}/N_{\rm H} = 1.086\int \pi a^2 Q_{\rm ext} (\lambda, a) \frac{1}{n_{\rm H}} \frac{dn}{da} da.
\end{equation}
Similarly, optical depths for scattering $\tau_{\lambda,{\rm sca}}$ and absorption $\tau_{\lambda,{\rm abs}}$ per $N_{\rm H}$ are calculated as
\begin{equation}
\tau_{\lambda,{\rm sca}}/N_{\rm H} = \int \pi a^2 Q_{\rm sca} (\lambda, a) \frac{1}{n_{\rm H}} \frac{dn}{da} da,
\end{equation}
\begin{equation}
\tau_{\lambda,{\rm abs}}/N_{\rm H} = \int \pi a^2 Q_{\rm abs} (\lambda, a) \frac{1}{n_{\rm H}} \frac{dn}{da} da.
\end{equation}
Then, albedo $\gamma_\lambda$ is expressed as
\begin{equation}
\gamma_\lambda = \frac{\int \pi a^2 Q_{\rm sca} (\lambda, a) \frac{1}{n_{\rm H}} \frac{dn}{da} da}{\int \pi a^2 Q_{\rm ext} (\lambda, a) \frac{1}{n_{\rm H}} \frac{dn}{da} da} = \frac{\tau_{\lambda,{\rm sca}}/N_{\rm H}}{\tau_{\lambda,{\rm ext}}/N_{\rm H}}.
\end{equation}

To calculate the DGL spectrum in a certain Galactic latitude $b$, scattering anisotropy should be taken into account because it is expected to cause $b$-dependence of the DGL (Jura 1979).
Scattering anisotropy of a single grain of its size $a$ is quantified as
\begin{equation}
g(\lambda,a) \equiv \int\Phi_{\lambda,a}(\theta)\cos\theta\, d\Omega,
\end{equation} 
where $\theta$, $\Phi_{\lambda,a}(\theta)$, and $\Omega$ are scattering angle, phase function and, solid angle, respectively.
The phase function is calculated by the Mie theory as functions of $\lambda$ and $a$.
By this definition, range of the quantity $g(\lambda,a)$ is $-1 \leq g(\lambda,a) \leq 1$.
Therefore, the $g$-factor of interstellar dust can be calculated as
\begin{equation}
g_\lambda = \frac{\int \pi a^2 Q_{\rm sca} (\lambda, a) g(\lambda, a) \frac{1}{n_{\rm H}} \frac{dn}{da} da}{\int \pi a^2 Q_{\rm sca} (\lambda, a) \frac{1}{n_{\rm H}} \frac{dn}{da} da}.
\end{equation}

If these quantities of interstellar dust and an ISRF spectrum $I_{\lambda, {\rm ISRF}}$ are supplied, DGL spectrum in an optically thin region can be represented as
\begin{equation}
I_{\lambda, {\rm DGL}} = \gamma_\lambda \tau_{\lambda, \rm{ext}} I_{\lambda, {\rm ISRF}} f(b,g_\lambda),
\end{equation}
where $f(b,g_\lambda)$ corresponds to an attenuation factor of the DGL according to the scattering anisotropy.
Jura (1979) assumes a plane-parallel galaxy with illuminating sources on the Galactic plane and approximate the factor $f(b,g_\lambda)$ as
\begin{equation}
f(b,g_\lambda) \approx 1-1.1g_\lambda\sqrt{\sin|b|}.
\end{equation}
In this approximation, they adopt an analytic phase function introduced by Henyey \& Greenstein (1941).
By using Equation (15), Equation (18) can be expressed as
\begin{equation}
I_{\lambda, {\rm DGL}} = \tau_{\lambda,{\rm sca}} I_{\lambda, {\rm ISRF}} f(b,g_\lambda).
\end{equation}

\subsection{Calculation of DGL spectra from dust models}

\begin{figure*}
\centering
 \includegraphics[width=10cm]{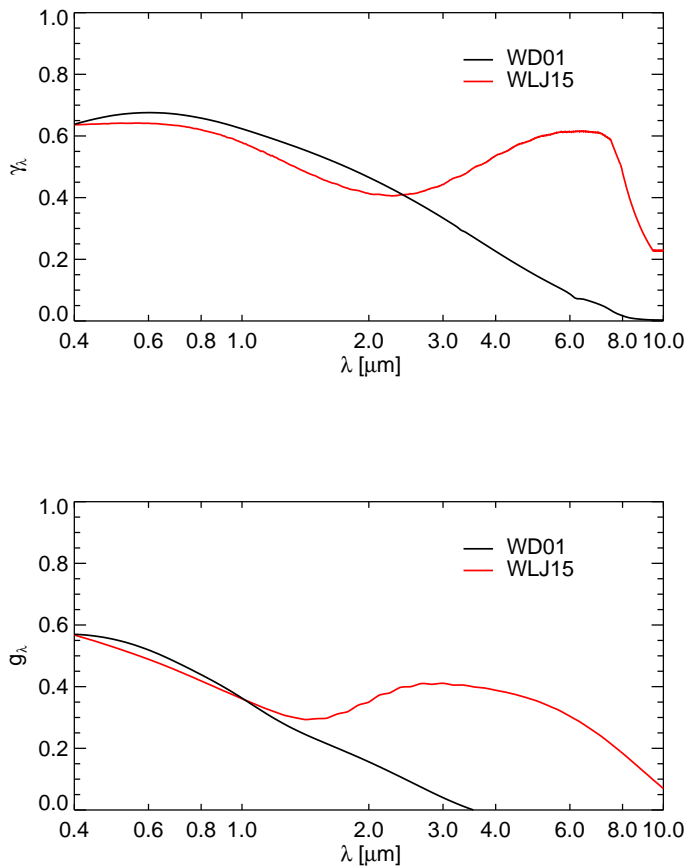}
 \caption{Spectra of albedo $\gamma_\lambda$ and $g$-factor $g_\lambda$ expected from the dust models of WD01 (black) and WLJ15 (red).
 The quantities of the WLJ15 model are calculated in the present study (Section 5.2), while those of WD01 are taken from Draine (2003a).}
\end{figure*}

Adapting the WLJ15 model to the general formulation described above, we calculate the DGL spectrum in the MBM32 field, where both optical and near-IR DGL results are available (Section 4.2).
The WLJ15 model is based on WD01, except for the VLGs composed of graphite.
According to Equation (2) in the paper of WLJ15, the VLG population in log-normal size distribution is given by
\begin{eqnarray}
\frac{1}{n_{\rm H}} \frac{dn}{da} &=& \frac{3}{(2\pi)^{3/2}} \times \frac{\exp(-4.5\sigma^2)}{\rho a_0^3 \sigma} \times \frac{b_{\rm VLG}\mu m_{\rm H}}{2} \nonumber \\ 
&\times& \frac{1}{a} \exp\Bigl[-\frac{1}{2}\Bigl\{\frac{\ln(a/a_0)}{\sigma}\Bigr\}^2\Bigr],
\end{eqnarray}
where $m_{\rm H}$ is mass of one hydrogen atom, $\rho$ and $\mu$ are mass density and molecular weight of carbon, respectively.
In WLJ15, this VLG population is added to the grain size distribution of the WD01 model and the parameters $b_{\rm VLG}$, $\sigma$, and $a_0$ are determined by fitting to the $R_V = 3.1$ extinction curve from Cardelli, Clayton, \& Mathis (1989).
The derived values are $b_{\rm VLG} \approx 137\,{\rm ppm}$, $\sigma \approx 0.3$, and $a_0 \approx 1.2\,\rm{\mu m}$.
By this fitting, size distribution of graphite and silicate grains smaller than sub-$\rm{\mu m}$ is also changed from WD01.
In addition, size distribution of graphite in WLJ15 shows a gap between the sub-$\rm{\mu m}$-sized grains and the VLGs (See Figure 2 of the WLJ15 paper).

In addition to the graphite and silicate grains, the WD01 model consists of PAH components represented as a sum of two log-normal distribution.
Their parameters are set to those adopted in Draine \& Li (2007; hereafter DL07) and the PAH components are assumed to have the same optical properties as graphite grains.

\begin{figure*}
\centering
 \includegraphics[width=15cm]{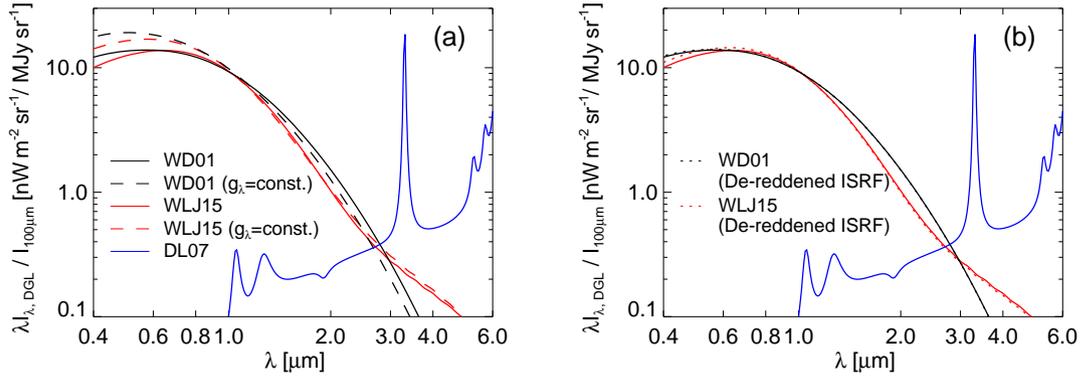}
 \caption{Comparison of the DGL model spectra.
 In both panels, black and red solid curves represent the DGL spectra calculated from the dust models of WD01 and WLJ15, respectively.
 In panel (a), black and red dashed curves are the models from WD01 and WLJ15, respectively, assuming no wavelength dependence of $g_{\lambda}$.
In panel (b), black and red dotted curves are the models from WD01 and WLJ15, respectively, assuming the de-reddened MMP83 spectrum for ISRF.
All of these spectra are scaled to the present observational result at $1.1\,\rm{\mu m}$. 
In both panels, the blue curve indicates intensity ratio of near-IR to $100\,\rm{\mu m}$ thermal emission, expected from the DL07 model with $U=1$ and $q_{\rm PAH} = 4.6\%$.
}
\end{figure*}

In the calculation of albedo $\gamma_\lambda$ and $g$-factor, the quantities $Q_{\rm sca} (\lambda, a)$, $Q_{\rm abs} (\lambda, a)$, and $g(\lambda,a)$ are taken from results of previous studies (Draine \& Lee 1984; Laor \& Draine 1993).
The data can be accessed through the website: ``www.astro.princeton.edu/$\sim$ draine/''.
Combining these values with the size distribution of WLJ15, we calculate the quantities $\gamma_{\lambda}$ and $g_{\lambda}$ from Equations (15) and (17), respectively.
Integral range of $a$ is from $0.001$ to $10\,\rm{\mu m}$ and this appears good approximation in the wavelengths of our interest.
Figure 11 compares our calculations of $\gamma_{\lambda}$ and $g_{\lambda}$ with the values from the WD01 dust (Draine 2003a).
Due to the VLG population, albedo and $g$-factor of the WLJ15 model show higher values from near- to mid-IR wavelengths ($\sim2$--$10\,\rm{\mu m}$), while these quantities from the WLJ15 model are comparable to those from WD01 in the shorter wavelengths ($\lesssim 1\,\rm{\mu m}$).

In the calculation of $f(b,g_\lambda)$, we adopt Equation (19) with the Galactic latitude of the MBM32 field, $b = 40.62^{\circ}$.
Sano \& Matsuura (2017) shows that the factor $f(b,g_\lambda)$ depends on assumed forms of phase function (Henyey \& Greenstein 1941; Draine 2003b) and the form (19) has large uncertainty.
To see the effect according to the uncertainty, we also calculate the DGL spectra in case of $g_\lambda = const.$, i.e., no $\lambda$-dependence in $f(b,g_\lambda)$.

As an ISRF spectrum, we adopt the model of Mathis, Mezger, and Panagia (1983; hereafter MMP83), which is based on observations of Galactic stars.
In this model, the ISRF spectrum in the solar neighborhood is expressed as a sum of four diluted blackbody spectra, corresponding to $10\,{\rm kpc}$ away from the Galactic center.
In addition to the original MMP83, we test de-reddening effect of ISRF because $I_{\lambda, {\rm ISRF}}$ in Equation (18) does not indicate the ISRF in the solar neibourhood but represent that in the field of MBM32 in reality.
For the similar reason, Brandt \& Draine (2012) also considers the de-reddening of ISRF in their DGL model assuming a plane-parallel galaxy.
In the present model assuming a dusty slab, the de-reddened ISRF spectrum is calculated as
\begin{equation}
I_{\lambda, {\rm ISRF}} = e^{\tau_{\lambda,{\rm ext}}} I_{\lambda, {\rm MMP83}}, 
\end{equation}
where $I_{\lambda, {\rm MMP83}}$ indicates the ISRF spectra of MMP83.
In general high-$b$ regions, $\tau_{\lambda,{\rm ext}}$ at $0.55\,\rm{\mu m}$ appears at most $\sim0.1$ according to the reddening map of SFD98 (e.g., Figure 11 of Brandt \& Draine 2012).
Therefore, we also calculate the DGL spectra in case of $\tau_{0.55\,\rm{\mu m},{\rm ext}} = 0.1$ to test the de-reddening effect of ISRF.

Using the ISRF spectra and dust properties of WLJ15, the DGL spectrum can be calculated from Equation (20).
For comparison, we also calculate the DGL spectrum from the WD01 model.
Figure 12 represents the DGL spectra calculated in the different situations described above.
All spectra are scaled to the present observation in the MBM32 field at $1.1\,\rm{\mu m}$.
In the panel (a), spectral difference between the cases of $g_\lambda$ and $g_\lambda = const.$ is less than a factor of 1.5 from optical to near-IR.
The panel (b) shows that the de-reddening effect of the ISRF has little influence to the DGL color from optical to near-IR.

To estimate possible contribution of the thermal dust emission in the near-IR, the model of DL07 is plotted in Figure 12.
The DL07 spectrum is represented as intensity ratio to the $100\,\rm{\mu m}$ emission.
In the DL07 model, ISRF scaling factor against MMP83 is set to $U=1$ and mass ratio of PAH to the total dust is assumed as $q_{\rm PAH} = 4.6\%$.
These parameters are the same as those adopted in the present calculation of the DGL spectra. 
As shown in Figure 12, DGL intensity is expected to exceed the DL07 model by one order of magnitude in $\lambda < 2\,\rm{\mu m}$, where the present study focuses on.
Thus, we neglect the thermal emission component in the further discussion.

\subsection{Implication of dust properties in the MBM32 field}

\begin{figure*}
\centering
 \includegraphics[width=15cm]{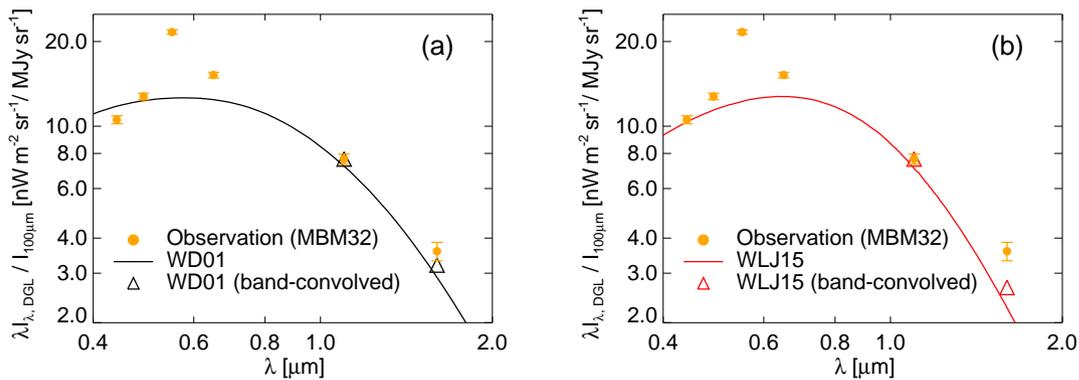}
 \caption{Comparison of our observation and DGL spectra calculated from the dust models of WD01 (panel a) and WLJ15 (panel b).
Orange circles represent intensity ratio of the optical and near-IR DGL to $100\,\rm{\mu m}$ emission in the MBM32 field.
Black and red solid curves indicate the DGL model spectra calculated from the WD01 and WLJ15 models, respectively.
At $1.1$ and $1.6\,\rm{\mu m}$, black and red triangles are the model values from WD01 and WLJ15, respectively, derived by convolving the spectra with the filter response of MIRIS (Figure 3).
In both panels, the model spectra are scaled so that the convolved value at $1.1\,\rm{\mu m}$ is identical to the observation.}
\end{figure*}

Figure 13 compares the DGL observed in MBM32 with the model spectra expected from the dust models of WD01 and WLJ15.
The model values in the $1.1$ and $1.6\,\rm{\mu m}$ bands are derived by convolving the spectra with the filter response of MIRIS (Figure 3).
The model spectra are scaled so that the convolved values are fitted to the observation at $1.1\,\rm{\mu m}$.
At $1.6\,\rm{\mu m}$, the observed value is closer to the WD01 model and is higher than the WLJ15 prediction by a factor of 1.5.
As shown in the figure, color of the near-IR DGL is closer to the model of WD01 and redder than the prediction from WLJ15.
From WD01 to WLJ15, the cutoff parameter of the sub-$\rm{\mu m}$-sized graphite changes from $0.428\,\rm{\mu m}$ to $0.173\,\rm{\mu m}$ due to the VLG component in the latter model.
Therefore, grains of high scattering efficiency in the near-IR should be fewer in the WLJ15 model.

In contrast to the situation in $\lambda =1$--$2\,\rm{\mu m}$, the WLJ15 model predicts redder DGL color in $\lambda > 3\,\rm{\mu m}$ (Figure 12).
However, it appears difficult to observe the DGL component in $\lambda > 3\,\rm{\mu m}$ because contributions of the thermal emission from PAH and very small grains become dominant in the longer wavelengths.
This is also suggested by several observations from near- to mid-IR (e.g., Flagey et al. 2006; Sano et al. 2016a; Sano \& Matsuura 2017).

If we adopt the WLJ15 size distribution in the MBM32 field, the near-IR ISRF spectrum should be redder to account for the observed DGL color.
In this case, the number of K- or M-type stars may increase to make the ISRF redder.
On the other hand, it appears more reasonable that $\rm{\mu m}$-sized VLGs are not present in the MBM32 field due to its low dust density ($A_V \lesssim 1\,{\rm mag}$).
According to numerical simulations of grain growth in the interstellar medium, the VLG formation can be enhanced in dense regions due to the coagulation process (Ormel et al. 2009; Hirashita \& Yan 2009; Hirashita \& Li 2013).
In several regions of dense molecular core ($A_V \gtrsim 10\,{\rm mag}$), scattered light component has been observed in the wavelength longer than $3\,\rm{\mu m}$ (Pagani et al. 2010; Steinacker et al. 2010; Steinacker et al. 2014).
These results can be interpreted as a piece of evidence of grain growth in the dense regions.
In addition, observations of the mid-IR extinction are limited to regions of enough extinction even in the mid-IR, such as sightlines toward Galactic plane.
In other words, the mid-IR flat extinction may not be applicable to optically thin diffuse clouds. 

Instead of graphite, Wang, Li, and Jiang (2015b) assumes VLGs composed of ${\rm H_2 O}$ ice to account for the surplus problem of oxygen in the interstellar medium (Whittet 2010) as well as the flat extinction curve in the mid-IR.
Though assumed size distribution of graphite and ${\rm H_2 O}$ ice is similar, light scattering properties can be different between these compositions.
It seems that the ${\mu m}$-sized ice grain model is promising since ice grains are known to be more reflective in the near-IR while graphite is more absoprtive.
As a future study, it would be useful to model the DGL spectrum by using the optical properties of ${\rm H_2 O}$ ice.

As shown in Figure 4 of WLJ15, VLGs mainly contribute to the thermal emission at $\lambda > 1000\,\rm{\mu m}$.
In the correlation analysis, therefore, using sub-mm data (e.g., Planck) instead of the $100\,\rm{\mu m}$ map may be more helpful to investigate the VLG population.
Such analysis should be conducted in a separate paper.

Both DGL spectra expected from WD01 and WLJ15 do not account for the excess observed at $0.55\,\rm{\mu m}$ (Figure 13).
This feature is sometimes referred to as extended red emission (ERE). 
As shown in Figure 8, possible existence of ERE in diffuse clouds has been reported so far (Guhathakurta \& Tyson 1989; Gordon et al. 1998; Witt et al. 2008; Matsuoka et al. 2011; Ienaka et al. 2013). 
The ERE could arise from the interaction of far-ultraviolet photons with ${\rm nm}$-sized interstellar dust (Witt \& Schild 1985; Darbon et al. 1999; Li \& Draine 2002; Li 2004; Witt et al. 2006). 
In contrast, DGL observation in wide-field blank sky does not suggest the presence of ERE (Brandt \& Draine 2012).
To constrain the origin of ERE, spectral study toward a number of clouds would be useful, though observations of diffuse light with high signal-to-noise ratio is difficult to achieve in general.

\section{Summary}

We study interstellar dust properties from DGL observations in three clouds at $1.1$ and $1.6\,\rm{\mu m}$ with wide-field space telescope MIRIS.
We searched for high-$b$ clouds of wide intensity range in far-IR $100\,\rm{\mu m}$ emission, which are appropriate for the DGL measurement.
Finally, we selected and observed the three regions including the field of MBM32 where optical DGL was also measured by Ienaka et al. (2013).

After the initial data reduction, we conduct flux calibration, point source masking, ZL removal, and smoothing of the near-IR images.
Then, we correlate the near-IR brightness with the SFD98 $100\,\rm{\mu m}$ intensity map.
We show linear correlations between the near-IR DGL and $100\,\rm{\mu m}$ emission.
The result is consistent with that of CIBER/LRS (Arai et al. 2015) and DIRBE (Sano et al. 2015). 
Owing to the wide-field space observation, our DGL results are the most precise among the literature.
In addition, we first derive the linear correlation between the near-IR and optical DGL in the field of MBM32.

In contrast to the conventional dust model composed of grains smaller than sub-$\rm{\mu m}$ (WD01), the new model including $\rm{\mu m}$-sized VLGs (WLJ15) succeeds in reproducing the flat shape of the mid-IR extinction curve observed in several fields of the Milky Way.
To investigate existence of the VLG population in the field of MBM32, we predict the DGL spectra from the two dust models and compare them with the observation.
In the near-IR, the observed DGL color is redder than the prediction of WLJ15 and closer to the WD01 model.
This result may indicate a redder ISRF spectrum in the MBM32 field or few VLGs in the region.
Since the observed field is optically thin, the latter case is consistent with some theoretical studies suggesting that grain growth proceeds in dense clouds.
In addition, our result indicates the presence of ERE in $\lambda \sim 0.6\,\rm{\mu m}$ as claimed by several observations so far.

\begin{ack}
We thank the referee for a number of comments that improved the manuscript.
This study is based on significant contributions from the MIRIS team. 
We are very grateful to the MIRIS staff for their observations and support in terms of data reduction. 
In addition, we appreciate a lot of advice on the target selection from Kazuhito Dobashi (Tokyo Gakugei University), Kimiaki Kawara (University of Tokyo), and Yoshiki Matsuoka (Ehime University). 
The authors acknowledge support from the Japanese Society for the Promotion of Science, KAKENHI (grant number 26800112, 21111004, 15H05744).
MIRIS team members at KASI acknowledge support from the Ministry of Science, ICT and Future Planning (MSIP) of Korea (NRF-2014M1A3A3A02034746).
\end{ack}





\end{document}